\documentclass[12pt,twoside]{article}

\usepackage{amsmath,amsfonts,amssymb,amsthm,mathabx}
\usepackage{times}
\usepackage{pdfsync}
\usepackage{cite}
\usepackage{url}
\usepackage{hyperref}
\usepackage{tensor}
\usepackage{color}
\usepackage{caption}
\usepackage{enumitem}
\usepackage{graphicx}
\usepackage{array}
\usepackage{amsbsy}
\usepackage{appendix}
\usepackage{mathrsfs}
\usepackage{physics}
\usepackage{cancel}
\usepackage{yfonts}
\usepackage{inputenc}
\usepackage{chngcntr}

\advance\textheight\topskip
\numberwithin{equation}{section} \makeatletter
\@addtoreset{equation}{section}

\newcommand{\D}{\text{d}}

\begin{document}

\title{Gravitons in a Casimir box}

\author{Francesco Alessio, Glenn Barnich, Martin Bronte}


\def\mytitle{Gravitons in a Casimir box}

\pagestyle{myheadings} \markboth{\textsc{\small F.~Alessio et al.}}
{\textsc{\small Massless spin 2 partition function with Casimir-type boundary conditions}}

\addtolength{\headsep}{4pt}

\begin{centering}

  \vspace{1cm}

  \textbf{\Large{\mytitle}}

  \vspace{1.5cm}

  {\large Francesco Alessio${}^{a,b}$, Glenn Barnich${}^{b}$, Martin
    Bonte$^{b}$}

\vspace{1cm}

\begin{minipage}{.9\textwidth}\small \it  \begin{center}
    ${}^a$ Dipartimento di Fisica ``E. Pancini'' and INFN Universit\`a
    degli studi di Napoli ``Federico II'', I-80125 Napoli, Italy
  \end{center}
\end{minipage}    

\vspace{1cm}

\begin{minipage}{.9\textwidth}\small \it  \begin{center}
  ${}^b$ Physique Th\'eorique et Math\'ematique, Universit\'e libre de
  Bruxelles and International Solvay Institutes, Campus
  Plaine C.P. 231, B-1050 Bruxelles, Belgium
\end{center}
\end{minipage}

\end{centering}

\vspace{1cm}
  
\begin{center}
  \begin{minipage}{.9\textwidth} \textsc{Abstract}. The partition function of
    gravitons with Casimir-type boundary conditions is worked out. The simplest
    box that allows one to achieve full analytical control consists of a slab
    geometry with two infinite parallel planes separated by a distance $d$. In
    this setting, linearized gravity, like electromagnetism, is equivalent to
    two free massless scalar fields, one with Dirichlet and one with Neumann
    boundary conditions, which in turn may be combined into a single massless
    scalar with periodic boundary conditions on an interval of length $2d$. When
    turning on a chemical potential for suitably adapted spin angular momentum,
    the partition function is modular covariant and expressed in terms of an
    Eisenstein series. It coincides with that for photons. At high temperature,
    the result provides in closed form all sub-leading finite-size corrections
    to the standard (gravitational) black body result. More interesting is the
    low-temperature/small distance expansion where the leading contribution to
    the partition function is linear in inverse temperature and given in terms
    of the Casimir energy of the system, whereas the leading contribution to the
    entropy is proportional to the area and originates from gravitons
    propagating parallel to the plates.
 \end{minipage}
\end{center}

\thispagestyle{empty}

\vfill
\newpage

\tableofcontents

\vfill
\newpage

\section{Introduction}
\label{sec:introduction}

What boundary conditions may be consistently imposed on gauge and gravitational
fields, both asymptotically or at finite distances, is an important problem that
has recently attracted a lot of attention because it is directly related not
only to asymptotic and residual gauge symmetries but also to the number and the
nature of the degrees of freedom of the system.

In his report on quantum fields in curved space (\cite{DeWitt:1975ys}, section
2.4), DeWitt explains that we know ``from years of experiment and years of model
building'' what a conductor is and what consistent boundary conditions to impose
for electric and magnetic fields. He then goes to infer meaningful boundary
conditions and boundary terms for the case of a massless scalar field.

In this work, we follow the spirit of DeWitt's approach and impose the analog of
``perfectly conducting'' boundary conditions on free massless spin 2 fields. We
will show that the partition function of gravitons in such a Casimir box can be
computed in a straightforward way and is identical to that of photons in the
original Casimir set-up, and also to that of a massless scalar field with
periodic boundary conditions in a box twice as big. This is not entirely
surprising since these are results on the level of free field theories. 

At this stage, we will limit ourselves to deriving the exact analytic and
modular covariant result that allows one to access the qualitatively very
different high and low temperature expansions, in agreement with recent results
on a Cardy formula and modular properties in higher dimensions
\cite{Cappelli:1988vw,Cardy:1991kr,Shaghoulian:2015kta}. We will not address
whether one may in principle confine gravitons to a box so as to achieve thermal
equilibrium
\cite{Smolin1984,garfinkle1985possibility,Dell1987,Padmanabhan_2003}. Neither
will we speculate on the physical meaning of perfect conductors for gravitons
nor comment on the relation to a recent study of the gravitational Casimir
effect at zero temperature with non-idealized boundary conditions
\cite{Quach:2015qwa}. As compared to the closely related computations in
\cite{Viaggiu:2017zax}, where a spherical box as a model for a Schwarzschild
black hole is considered, the use of a slab geometry simplifies a detailed
discussion of the contribution of various degrees of freedom to the partition
function and thus to the entropy, in much the same way than such a geometry
makes computations and conceptual issues relatively straightforward in the
original Casimir analysis \cite{casimir1948attraction}, whereas the spherically
symmetric case \cite{Boyer:1968uf,doi:10.1063/1.1665021} is much more involved.
Finally, we will not discuss here of how such computations in linearized gravity
could be relevant for black hole physics in full non-linear general relativity
or compare to results in the context of cosmological models
\cite{Brandenberger:1992jh}.

The paper is organized as follows. As a useful preparation, we re-discuss in
appendix \ref{sec:reduced-phase-space-2}, our analysis
\cite{Barnich:2018zdg,Barnich2019} of photons with Casimir boundary conditions
by using polarization vectors adapted to $E$ and $H$ modes. We also briefly
review in appendix \ref{sec:reduced-phase-space} the well-known reduced phase
space of linearized gravity in empty space. In the main text, we start in
section \ref{sec:casim-bound-cond}, by defining Casimir-type boundary conditions
with the associated mode expansions for metric perturbations and their canonical
momenta. This allows us in section \ref{sec:reduced-phase-space-1} to work out
the physical degrees of freedom in this precise set-up and to isolate the sector
with vanishing transverse momentum, or in other words, the sector of
gravitational waves propagating parallel to the plates. The next step in section
\ref{sec:scal-field-form} consists in organizing all physical degrees of
freedom, first in terms of two massless scalar fields with Dirichlet and Neumann
boundary conditions, respectively and then in terms of a single scalar field. In
order to do so, we follow recent work on the finite temperature electromagnetic
Casimir effect and modular invariance in this context \cite{Alessio:2020okv} to
show that this scalar field should satisfy periodic boundary on an interval of
double the length of the separation of the plates. This allows us in section
\ref{sec:extend-part-funct} to simply infer the exact result for the partition
function for gravitons, with a chemical potential for suitably modified spin
angular momentum turned on, from the well-studied case of a massless scalar
field with linear momentum in the compact direction turned on
\cite{Cappelli:1988vw,Cardy:1991kr,Shaghoulian:2015kta,Alessio2020}, together
with modular properties and consistent high and low temperature expansions.

Arguably the most interesting aspect of the exact analytic result is the fact
that the low temperature/small box expansion is qualitatively very different
from the more standard high temperature/large volume expansion. The latter is
dominated by the standard black body result, up to finite size corrections. At
low temperature however, the leading contribution to the partition function
scales like inverse temperature and is responsible for the zero temperature
gravitational Casimir force. It does not contribute to the entropy. The
sub-leading contribution to the partition function is then the leading
contribution to the entropy. It is proportional to the area of the plates and
our analysis shows that it originates from gravitons that propagate parallel to
those plates. The dynamics of those gravitons is equivalent to that of photons
in two spatial dimensions, which in turn is described by a massless scalar field
in 2 spatial dimensions. Let us stress here that these degrees of freedom are
due to the non-trivial boundary conditions imposed on the fields but, except for
the particle present in the spectrum, there is ultimately no interpretation in
terms of residual gauge degrees of freedom nor a direct relation to large gauge
transformations, contrary to what is implied in
\cite{Barnich:2018zdg,Barnich2019}.

\section{Casimir boundary conditions and mode expansions}
\label{sec:casim-bound-cond}

We want to promote the empty space analysis of massless spin 2 fields in the
Hamiltonian approach, as reviewed in appendix \ref{sec:reduced-phase-space}, to
a slab geometry with two infinite parallel planes at $x^3=-\frac{d}{2}$ and
$x^3=\frac{d}{2}$. Let $a=1,2$, $i=a,3$, $V=L_1L_2d$ the volume of the system
with $L_a$ taken large in the end, and $d$ the separation of the plates. Let
$k_3=\frac{\pi n_3}{d}$, $k_a=\frac{2\pi n_a}{L_a}$ (with no summation over $a$).

In analogy with the electromagnetic case reviewed in appendix
\ref{sec:reduced-phase-space-2}, we define Casimir-type boundary conditions by
requiring that $h_{ab},\pi^{ab}$ and $h_{33},\pi^{33}$ satisfy Dirichlet
conditions, while $h_{a3},\pi^{a3}$ satisfy Neumann conditions. More explicitly,
at the boundary $x^3=-\frac{d}{2}$ and $x^3=\frac{d}{2}$, we require
\begin{equation}
  \label{eq:78}
  h_{ab}=0=\pi^{ab},\quad h_{33}=0=\pi^{33},
  \quad (\partial_3 h_{a3})=0=(\partial_3 \pi^{a3}). 
\end{equation}
Let us use $T^{ij}$ for either $h_{ij}$ or $\pi^{ij}$ and define, as in
\eqref{eq:14}, a complete set of scalar functions $\psi^H_k$ with $n_a,n_3>0$
adapted to Dirichlet boundary conditions, and $\psi^E_k$ with $n_a,n_3\geq 0$
adapted to Neumann boundary conditions.

The boundary conditions on the canonically conjugate variables are thus
implemented through the mode expansions \footnote{The factor $(-i)$ in front of
  the expansion of $(h_{a3},\pi^{a3})$ is chosen for later convenience.}
\begin{equation}
  \label{eq:2a}
  T^{ab}(x)=\sum_{n_a,n_3>0}T^{ab}_{k}\psi^H_k,\quad
  T^{33}(x)=\sum_{n_a,n_3>0}T^{33}_{k}\psi^H_{k},\quad
  T^{a3}(x)=(-i)\sum_{n_a,n_3\geq 0}T^{a3}_{k}\psi^E_{k},
\end{equation}
where we take $T^{ab}_{k_a,0}=0=T^{33}_{k_a,0}$. Reality conditions
are
\begin{equation}
  \label{eq:3a}
  T^{ab}_{k_a,k_3}=T^{*ab}_{-k_a,k_3},\quad T^{33}_{k_a,k_3}=T^{*33}_{-k_a,k_3},\quad
  T^{a3}_{k_a,k_3}=-T^{*a3}_{-k_a,k_3},
\end{equation}
while parity conditions are
\begin{equation}
  \label{eq:79}
  T^{ab}_{k_a,k_3}=-T^{ab}_{k_a,-k_3},\quad T^{33}_{k_a,k_3}=-T^{33}_{k_a,-k_3},\quad
  T^{a3}_{k_a,k_3}=T^{a3}_{k_a,-k_3}. 
\end{equation}
Odd variables do not have modes at $n_i=0$, $T^{ab}_{0,0}=0=T^{33}_{0,0}$, but
there are two such modes in the even sector, $T^{a3}_{0,0}\neq 0$.

Note that these boundary conditions have been constructed in such a way that the
left hand sides of the constraints have well defined expansions, either in terms
of sines or in terms of cosines. If one requires the lapse and the shift to
impose all constraints in a first order variational principle and not to
introduce spurious variables, it follows from taking into account the mode
expansion of the constraints \eqref{eq:57}, that the lapse $n$ and $n^a$ should
satisfy Dirichlet while $n^3$ should satisfy Neumann conditions,
\begin{equation}
  \label{eq:103}
  n=\sum_{n_a,n_3>0}n_k\psi_k^H,\quad n^a=\sum_{n_a,n_3>0}n^a_k\psi_k^H,\quad
  n^3=\sideset{}{'}\sum_{n_a,n_3\geq 0}n^3_k\psi_k^E, 
\end{equation}
so that $n,n^a$ are odd, while $n^3$ is even under $k_3\to -k_3$.

\section{Reduced phase space sectors}
\label{sec:reduced-phase-space-1}

In terms of these decomposition,
the kinetic term is given by
\begin{equation}
  \label{eq:81}
  \int \D^3x\ \partial_0 h_{ij}\pi^{ij}=\sum_{n_a,n_3\geq 0}\partial_0
  h_{ij}\pi^{*ij},
\end{equation}
where we sometimes omit the $k$ dependence of the Fourier coefficients for
notational simplicity. The Pauli-Fierz Hamiltonian \eqref{eq:56} is given by
\begin{equation}
  \label{eq:80}
  H_{\rm PF}=\sum_{n_a,n_3\geq 0}\Big[\pi^{ij}\pi^*_{ij}-\frac{1}{2}\pi\pi^*
  +\frac{1}{4}k^2 h_{ij}h^{*ij}-\frac{1}{2}k_i h^{ij}k^lh_{lj}^*+\frac{1}{2}k^ik^jh h_{ij}^*
  -\frac{1}{4}k^2 hh^*\Big]. 
\end{equation}

\subsection{Particle sector}
\label{sec:particle-sector}

We start with the particle sector, which has to be treated separately since the
change of variables that we will use below does not apply to this sector. For
$n_i=0$, we define $h_{a3,0,0}=\sqrt{2}q^a$ and $\pi^{a3}_{0,0}=\frac{1}{\sqrt
  2}p_a$. From the expansion of the kinetic term, these variables are
canonically conjugate. None of these variables is involved in the constraints nor
affected by a gauge transformation generated by these constraints. Their
contribution to the Pauli-Fierz Hamiltonian is that of two particles of unit
mass,
\begin{equation}
  \label{eq:82}
  H_{n_i=0}=\frac{1}{2}p^ap_a. 
\end{equation}

\subsection{Adapted polarization tensors}
\label{sec:adapt-polar-tens}

For $n_i\neq 0$, one may use the polarization tensors as defined in
\eqref{eq:63}-\eqref{eq:66} to decompose the components. Furthermore, it will
turn out to be convenient to use the polarization tensors built out of the
adapted polarization vectors ${e_\lambda}^i$, $\lambda=(H,E)$ given in
\eqref{eq:9} rather than to use generic transverse vectors ${e_\alpha}^i$.

As a consequence,
\begin{equation}
  \label{eq:96} \left\{ \begin{array}{ll} {e_{TT+}}^{ab}=\frac{1}{\sqrt
2k_\perp^2k^2}{(k^2\epsilon^{ac}k_c\epsilon^{bd}k_d-k^2_3k^ak^b)}, &
{e_{TT\times}}^{ab}=\frac{k_3}{\sqrt
2k_\perp^2k}{(\epsilon^{ac}k_ck^b+k^a\epsilon^{bc}k_c)},\\
{e_{TT+}}^{a3}=\frac{1}{\sqrt 2k^2}k^ak^3,& {e_{TT\times}}^{a3}=-\frac{1}{\sqrt
2k}\epsilon^{ac}k_c,\\ {e_{TT+}}^{33}=-\frac{1}{\sqrt 2k^2}k_\perp^2, &
{e_{TT\times}}^{33}=0.
\end{array}\right.
\end{equation}
All variables now have the same reality conditions
\begin{equation}
  \label{eq:3a}
  T^{TTs}_{k_a,k_3}=T^{*TTs}_{-k_a,k_3},\quad T^{T}_{k_a,k_3}=T^{*T}_{-k_a,k_3},\quad
  T^{LT\lambda}_{k_a,k_3}=T^{*LT\lambda}_{-k_a,k_3},\quad T^{LL}_{k_a,k_3}=T^{*LL}_{-k_a,k_3}, 
\end{equation}
while parity conditions become
\begin{equation}
  \label{eq:79}
  \begin{split}
    & T^{TT\times}_{k_a,k_3}=T^{TT\times}_{k_a,-k_3},\quad T^{LTE}_{k_a,k_3}=T^{LTE}_{k_a,-k_3},\\
    & T^{TT+}_{k_a,k_3}=-T^{TT+}_{k_a,-k_3},\quad
    T^{T}_{k_a,k_3}=-T^{T}_{k_a,-k_3},\quad
    T^{LTH}_{k_a,k_3}=-T^{LTH}_{k_a,-k_3},\quad 
    T^{LL}_{k_a,k_3}=-T^{LL}_{k_a,-k_3}.
\end{split}
\end{equation}

If $\Xi=(TTs,T,LT\lambda,LL)$, the kinetic term is
\begin{equation}
  \label{eq:84}
  \int \D^3x\ \partial_0 h_{ij}\pi^{ij=}\sideset{}{'}\sum_{n_a,n_3\geq 0}\ \partial_0h_{\Xi}\pi^{*\Xi},
\end{equation}
so that the Poisson brackets are
\begin{equation}
  \label{eq:86}
  \{h_{\Xi k},\pi^{*\Gamma}_{k'}\}=\delta^\Gamma_\Xi\prod_{i=1}^3\delta_{n_i,n'_i}. 
\end{equation}
The Hamiltonian constraints \eqref{eq:57} become
\begin{equation}
  \label{eq:83}
  \begin{split}
    \mathcal H_a&=-2i\sum_{n_a,n_3>0}[k^b\pi_{ab}+k^3\pi_{a3}]{\psi^H_k}
    =-2i\sum_{n_a,n_3>0}k[{e^\lambda}_a\frac{\pi_{LT
      \lambda}}{\sqrt 2}
    +{e^\parallel}_a\pi_{LL}]{\psi^H_k},\\
    \mathcal H_3&=(-2)\sideset{}{'}\sum_{n_a,n_3\geq 0}
    [k^b\pi_{b3}+k^3\pi_{33}]\psi^E_k
    =-2\sideset{}{'}\sum_{n_a,n_3\geq 0}
    k[{e^\lambda}_3\frac{\pi_{LT \lambda}}{\sqrt 2}+{e^\parallel}_3\pi_{LL}]\psi^E_k, \\
    \mathcal H_{\perp}&=-\sum_{n_a,n_3>0}k^2 \sqrt 2  h_T\psi^H_k.
\end{split}
\end{equation}
The Pauli-Fierz Hamiltonian \eqref{eq:56} is given by 
\begin{equation}
  \label{eq:91}
  H_{\rm PF}=\sideset{}{'}\sum_{n_a,n_3\geq 0}\big[\pi^{TTs}\pi^*_{TTs}+
  \pi^{LT\lambda}\pi^*_{LT\lambda}+\frac{1}{2}\pi^{LL}\pi^*_{LL}
  +\frac{1}{4}k^2(h_{TTs}h^{*TTs}-h_Th^{*T})\big]. 
\end{equation}

\subsection{Modes propagating parallel to the plates}
\label{sec:massl-scal-field}

For $n_i\neq 0$, let us start the analysis with the sector at $n_3=0$. It is
useful to consider the dyad
\begin{equation}
  {e_\perp}^b(k_a,0)=\frac{\epsilon^{bc}k_c}{k_\perp},\quad {e_\parallel}^c(k_a,0)
  =\frac{k^c}{k_\perp}, \label{eq:88}
\end{equation}
in terms of which a two-dimensional vector $T^b_{k_a,0}$ may be decomposed as
\begin{equation}
T^b_{k_a,0}=T^\perp_{k_a,0}{e_\perp}^b(k_a,0)+T^\parallel_{k_a,0} {e_\parallel}^b(k_a,0).\label{eq:92}
\end{equation}

When using $T^{ab}_{k_a,0}=0=T^{33}_{k_a,0}$ with only $T^{a3}_{k_a,0}\neq 0$,
it is straightforward to check using the explicit form of the polarization
vectors that
\begin{equation}
  \label{eq:87}
  \begin{split}
    T^{LL}_{k_a,0}=0,\quad T^T_{k_a,0}=0,\quad T^{LTH}_{k_a,0}=0, \quad T^{TT+}_{k_a,0}=0,\\
    T^{LTE}_{k_a,0}= -\frac{2}{\sqrt 2} T^{\parallel 3}_{k_a,0},\quad
    T^{TT\times}_{k_a,0}=-\frac{2}{\sqrt 2} T^{\perp 3}_{k_a,0}.
\end{split}
\end{equation}

At $n_3=0$, $n_a\neq 0$, the constraints reduce to 
\begin{equation}
\pi^{LTE}_{k_a,0}=0,\label{eq:104}
\end{equation}
or equivalently to $\pi^{\parallel 3}_{k_a,0}=0$. On the constraint surface, the
contribution from the $n_3=0$ modes to the first order action Pauli-Fierz action
then reduces to the contribution of the canonical pair
\begin{equation}
h^{TT\times}_{k_a,0}=-\frac{2}{\sqrt
  2}h^{\perp 3}_{k_a,0},\quad \pi^{TT\times}_{k_a,0}
=-\frac{2}{\sqrt 2}\pi^{\perp 3}_{k_a,0},\label{eq:93}
\end{equation}
with Hamiltonian
\begin{equation}
  \label{eq:90}
  H_{{\rm PF},n_3=0}\approx \sideset{}{'}\sum_{n_a}(\pi^{TT\times}_{k_a,0}\pi^{*TT\times}_{k_a,0}
  +\frac{1}{4}k_\perp^2 h^{TT\times}_{k_a,0}h^{*TT\times}_{k_a,0}). 
\end{equation}

After the canonical transformation $\pi^{TT\times}_{k_a,0}\to \frac{1}{\sqrt
  2}\pi^{TT\times}_{k_a,0}$, $h^{TT\times}_{k_a,0}\to \frac{1}{\sqrt
  2}h^{TT\times}_{k_a,0}$, it is obvious that this collection of harmonic
oscillators is the same than one would get from the expansion of a massless
scalar field in $2$ spatial dimensions with periodic boundary conditions
(omitting its particle mode). That is the reason why we sometimes refer to this
sector as the lower dimensional scalar field. This sector is somewhat unexpected
when comparing to the empty space analysis in appendix
\ref{sec:reduced-phase-space} and is due to the perfectly conducting boundary
conditions discussed in the previous section. From the explicit discussion of
the structure of the constraints and the Hamiltonian, we have thus shown the
following:

{\em With Casimir-type boundary conditions \eqref{eq:78}, gravitational waves
  propagating parallel to the plates are equivalently described by
  electromagnetic waves in $2$ spatial dimensions, whose physical degrees of
  freedom are in turn described by a massless scalar field in $2$ spatial
  dimensions.}

\subsection{Modes with a normal component of propagation}
\label{sec:modes-prop-norm}

When $n_3\neq 0$, the constraints $\mathcal H_i=0$ in \eqref{eq:83} are equivalent
to
\begin{equation}
  \label{eq:95}
  {e^\lambda}_i\frac{\pi_{LT\lambda}}{\sqrt 2}+{e^\parallel}_i \pi_{LL}=0. 
\end{equation}
After contracting with ${e_A}^i$, the constraints \eqref{eq:83} are then 
equivalent to the vanishing of specific components, 
\begin{equation}
  \label{eq:94}
\pi_{LT\lambda}=0=\pi_{LL},\quad h_{T}=0.
\end{equation}
On the constraint surface, the only variables that remain in the first order
Hamiltonian action are $h_{TTs},\pi^{TTs}$ with
\begin{equation}
  \label{eq:89}
  H_{{\rm PF},n_3>0}\approx \sum_{n_a,n_3> 0}(\pi_{TTs}\pi^{*TTs}
  +\frac{1}{4}k^2 h_{TTs}h^{*TTs}).
\end{equation}

Note that, if desired, one may also work with the gauge fixing conditions
\begin{equation}
\triangle \pi-\partial_i\partial_j\pi^{ij}=0,\quad -2\partial_jh^{ij}=0.\label{eq:97}
\end{equation}
and consider Dirac brackets. 

\section{Scalar field formulations}
\label{sec:scal-field-form}

\subsection{Bromwich-Borgnis fields}
\label{sec:bromw-borgn-fields}

Disregarding particle zero modes, the reduced phase space consists of two scalar
fields and their momenta, one pair satisfying Dirichlet and the other Neumann
boundary conditions,
\begin{equation}
  \label{eq:99}
  \begin{split}
  h^H=\sum_{n_a,n_3> 0}h^{TT+}_{k}\psi^H_k,\quad \pi^H=\sum_{n_a,n_3> 0}\pi^{TT+}_{k}\psi^H_k,\\
  h^E=\sideset{}{'}\sum_{n_a,n_3\geq 0}h^{TT\times}_k\psi^E_k,\quad 
  \pi^E=\sideset{}{'}\sum_{n_a,n_3\geq 0}\pi^{TT\times}_k\psi^E_k.
\end{split}
\end{equation}
On account of the form of the Pauli-Fierz Hamiltonian, \eqref{eq:89} and
\eqref{eq:90}, there is thus a direct gravitational analogue of the electromagnetic results
reviewed in appendix \ref{sec:borgnis-fields}: 

{\em With Casimir-type boundary conditions \eqref{eq:78} the dynamics of
  physical gravitons is equivalent to that of two massless scalar fields, one
  with Dirichlet and one with Neumann boundary conditions, with action
\begin{equation}
  \label{eq:100}
  S[h^\lambda]=-\frac{1}{2}\int\D^4 x\ \partial_\mu h^\lambda\partial^\mu h_\lambda,\quad \lambda=(H,E). 
\end{equation}
}

Note that the massless scalar field in $2$ spatial dimensions discussed in
section \ref{sec:massl-scal-field} corresponds to the modes at $k_3=0$ contained
in $h_E$.

The set-up follows closely the electromagnetic case discussed in appendix
\ref{sec:reduced-phase-space-2}, with the following substitutions:
\begin{equation}
  \label{eq:107}
  A^E_k\to \frac{1}{\sqrt{2}} h^{TT\times}_k,\quad \pi^E_k\to \sqrt 2  \pi ^{TT\times}_k,\quad
  A^H_k\to \frac{1}{\sqrt{2}} h^{TT+}_k,\quad \pi^H_k\to \sqrt 2  \pi ^{TT+}_k.
\end{equation}
This defines the oscillator variables $a^\lambda_k,a^{*\lambda}_k$ in
\eqref{eq:13} in terms of the Fourier components of the physical gravitational
fluctuations with standard Poisson brackets \eqref{eq:23}. The Pauli-Fierz
Hamiltonian without the particle contributions, given by the sum of the pieces
in \eqref{eq:90} and of \eqref{eq:89}, is equivalent to
\begin{equation}
  \label{eq:105}
  H'_{\rm PF}=\sideset{}{'}\sum_{\lambda,n_a,n_3\geq 0}
  \frac{1}{2} (\pi^\lambda_{k}\pi^{*\lambda}_k+
  k^2h_k^\lambda h^{*\lambda}_k)
  =\sideset{}{'}\sum_{\lambda,n_a,n_3\geq
    0}\frac{k}{2} (a^\lambda_ka^{*\lambda}_k+ a^{*\lambda}_ka^\lambda_k).
\end{equation}

{\bf Remark:}

In momentum space, the shift can be further expanded as
$n^i_k=n^\lambda_k{e_\lambda}^i+n^\parallel_k {e_\parallel}^i$, with
$n^H_k,n^\parallel_k$ even and $n^E_k$ odd. In a variational principle with
constraints included, variation with respect to these variables then directly
impose the constraints \eqref{eq:104} and \eqref{eq:94}. These variables may be
lumped into fields
\begin{equation}
  \label{eq:101}
  n^{H,\parallel}=\sum_{n_a,n_3> 0}n^{H,\parallel}_{k}\psi^H_k,\quad
  n^{E}=\sideset{}{'}\sum_{n_a,n_3\geq 0}n^{E}_{k}\psi^E_k,  
\end{equation}
together with the expansion of the shift in \eqref{eq:103}.

In a Hamiltonian BFV-BRST approach, spatial diffeomorphism ghosts and the ghost
associated to deformations normal to the surface that goes with the Hamiltonian
constraint follow the same pattern, with the same expansion holding for their
momenta. In an extended phase space, there are further canonically conjugate
pairs of scalar fields,
\begin{equation}
  \label{eq:102}
  \begin{split}
    & \phi^{T,LTH,LL} = \sum_{n_a,n_3> 0}h^{T,LTH,LL}_{k}\psi^H_k,\quad
    \pi^{T,LTH,LL} = \sum_{n_a,n_3> 0}\pi^{T,LTH,LL}_{k}\psi^H_k,\\
    & \phi^{LTE} = \sideset{}{'}\sum_{n_a,n_3\geq 0}h^{LTE}_{k}\psi^E_k,\quad
    \pi^{LTE} = \sideset{}{'}\sum_{n_a,n_3\geq 0}\pi^{LTE}_{k}\psi^E_k.
  \end{split}
\end{equation}
Quantization of all these variables should be done in terms of quartets so as to
have achieve equivalence with the reduced phase space quantization described
above. After integration over all momenta in the BRST gauge fixed path integral,
this then gives rise to the BRST gauge fixed path integral in the Lagrangian
approach, including the correct boundary conditions on all fields. Such an
extended setting is useful in order to set up computations in covariant gauges
(see e.g.~\cite{Henneaux:1992ig}) for a review), and also when including
interactions. We will not pursue this further here, but focus on reduced phase
space quantization.

\subsection{Additional observable}
\label{sec:addit-observ}

In order to achieve full modular covariance of the partition function below, an
additional observable is of interest. Its expression is most transparent in
terms of generalized vector calculus operations that feature prominently in the
context of the Hamiltonian approach to duality invariance
\cite{Deser:1976iy,Henneaux:2004jw,Deser:2004xt,Deser:2005sz,Barnich:2008ts}.
First, one considers the generalized vector product between a vector $v^k$ and a
symmetric tensor $T_{ij}$
\begin{equation}
  \label{eq:109}
  (\vec v\times T)_{ij}=\frac{1}{2}(\epsilon_{ilm}v^lT^m_j+\epsilon_{jlm}v^lT^m_i).
\end{equation}
This is then implies that the generalized curl for a symmetric spacetime tensor is,
\begin{equation}
  \label{eq:108}
  (\mathcal O T)_{ij}=(\vec \nabla\times T)_{ij}=\frac{1}{2}(\epsilon_{ilm}\partial^lT^m_j+\epsilon_{jlm}
  \partial^lT^m_i).
\end{equation}
It rotates electric and magnetic components and projects out both the trace part
$(T)$ and the longitudinal part of the longitudinal part $(LL)$. Then there is
the operator that projects out the transverse components of the longitudinal
part $(LT)$,
\begin{equation}
  (\mathcal P T)_{ij}=-\triangle T_{ij}+\partial_i \partial_m T^m_j
  +\partial_j \partial_m T^m_i. 
\label{eq:117}
\end{equation}
Combining these two operations gives a projector onto the transverse-traceless
part,
\begin{equation}
  \label{eq:118}
  (\mathcal P^{TT} T)_{ij}=-(\mathcal O \mathcal P T)_{ij}=-(\mathcal P \mathcal O T)_{ij}. 
\end{equation}
Details on how these operator act are provided in appendix
\ref{sec:gener-curl-proj}.

For simplicity, we start by working in the reduced phase space with
$T^T_{k}=T^{LT\lambda}_k=T^{LL}_k=0$. In terms of space-time fields, we thus
define $T_{TT}^{ij}$ through
\begin{equation}
  \label{eq:116}
  \begin{split}
    &  T^{ab}_{TT}=\sum_{n_a,n_3>0}{e_{TTs}}^{ab}\,T^{TTs}_k\psi_k^H,\quad
    T^{33}_{TT}=\sum_{n_a,n_3>0}{e_{TTs}}^{ab}\,T^{TTs}_k\psi_k^H,\\
    & T^{a3}_{TT}=-i\sideset{}{'}\sum_{n_a,n_3\geq 0}{e_{TTs}}^{a3}\,T^{TTs}_k\psi_k^E.
  \end{split}
\end{equation}

In analogy with \eqref{eq:25}, we then consider spin angular momentum in linearized
gravity, defined on the reduced phase space by
\begin{equation}
  \label{eq:98}
  J_i=\int_V\D^3 x\ \pi^{mn}_{TT}(\vec e_i\times h^{TT})_{mn}, 
\end{equation}
where $\vec e_i$ is the unit vector along $x^i$ in position space. When using
that
\begin{equation}
  \label{eq:110}
  {e_{TT+}}^{ij}{e_{TT\times}}^{mn}-{e_{TT\times}}^{ij}{e_{TT+}}^{mn}=
  \frac{1}{2}\big[\epsilon^{inl}(\delta^{jm}-\frac{k^jk^m}{k^2})
  +\epsilon^{jml}\left(\delta^{in}-\frac{k^ik^n}{k^2}\right)\big]\frac{k_{l}}{k},
\end{equation}
which follows from \eqref{eq:106}, one finds in terms of modes that 
\begin{equation}
  \label{eq:111}
 J_3=\sum_{n_a,n_3>0}\frac{k_3}{k}(h^{TT+}_k\pi^{*TT\times}_k-h^{TT\times}_k\pi^{*TT+}_k)
    = \sum_{n_a,n_3>0}\frac{ik_3}{k}(a^{*H}_ka^{E}_k-a^{*E}_ka^{H}_k). 
\end{equation}

The observable we are interested in is
\begin{equation}
  \label{eq:112}
  P_3^S=-\int_V \D^3x\ \partial_3(\mathcal O h)_{mn}
  \frac{1}{\sqrt{-\triangle}^3}(\mathcal P\pi)^{mn}, 
\end{equation}
which is gauge invariant without the need to put unphysical components to zero
by hand. When using the actions of $\mathcal O$ and $\mathcal P$ established in
appendix \ref{sec:gener-curl-proj}, together with the orthonormality properties
\eqref{eq:67}, it is straightforward to verify that its mode expansion differs
from that of $J_3$ by multiplying each term in momentum space by $k$,
\begin{equation}
  \label{eq:113}
  P^S_3=\sum_{n_a,n_3>0}{k_3}(h^{TT+}_k\pi^{*TT\times}_k-h^{TT\times}_k\pi^{*TT+}_k)
    = \sum_{n_a,n_3>0}{ik_3}(a^{*H}_ka^{E}_k-a^{*E}_ka^{H}_k). 
\end{equation}

\subsection{Single scalar field formulation}
\label{sec:single-scalar-field-1}

In the slab geometry considered here, one can go one step further and combine
the Fourier components of the two scalar fields with Neumann and Dirichlet
boundary into those of a single massless free scalar field with periodic
boundary conditions on an interval of length $2d$.

More precisely, from the expression of the Pauli-Fierz Hamiltonian
\eqref{eq:105} and the observable $P_3^S$ \eqref{eq:113} in terms of modes and
oscillators, one may show directly that these coincide with the standard
Hamiltonian, omitting its particle mode, and linear momentum in the $x^3$
direction, of a massless scalar field with periodic boundary conditions in a
rectangular box with sides $L_1,L_2,2d$. The proof is identical to that in the
electromagnetic case reviewed in appendix \ref{sec:single-scalar-field} when
taking into account the identifications \eqref{eq:107}. In summary,

{\em With Casimir-type boundary conditions \eqref{eq:78}, the dynamics of
  physical gravitons is equivalent to that of a massless scalar field with two
  large dimensions and with periodic boundary conditions on an interval of
  length $2d$.}

\section{Partition function}
\label{sec:extend-part-funct}

From the above equivalence, the partition function for gravitons with
Casimir-type boundary conditions may be obtained from that of a massless scalar
field on $\mathbb S^1_{2d}\times \mathbb R^2$. In turn, as shown in
\cite{Alessio:2020okv} and reviewed in appendix \ref{sec:reduced-phase-space-2},
the same equivalence holds in the electromagnetic case. In the absence of a
chemical potential, $\alpha=0$, up to a subtlety related to whether the black
body result is subtracted or not, the result can thus also be obtained from the
literature on the finite temperature electromagnetic Casimir effect
\cite{Mehra:1967wf,Brown:1969na} (see also
\cite{Dowker_1976,BALIAN1978165,10.1143/PTP.75.262,Ambjorn:1981xw,Ambjorn:1981xv,%
  Plunien:1987fr,Lutken:1988ge,Ford:1988gt,PhysRevD.40.4191} and
\cite{Plunien:1986ca,Nesterenko:2005xv,Bordag:2009zzd}) for reviews). It is
however more economical to derive the result, whether the chemical potential is
turned on or not, directly from the scalar field formulation. With $\alpha\neq
0$, this has been discussed instance in \cite{Cappelli:1988vw}. The detailed
derivation by a variety of different methods of the expressions provided below
can be found in \cite{Alessio2020}.

\subsection{Real analytic Eisenstein series from functional approach}
\label{sec:expr-terms-analyt}

The partition function
\begin{equation}
  \label{eq:119}
  Z(\beta,\alpha)={\rm Tr}\ e^{-\beta \hat H+i\alpha \hat P^S_3}, 
\end{equation}
may be conveniently expressed in terms of the modular
parameter defined by
\begin{equation}
  \label{eq:37}
  \tau=\frac{\alpha+i\beta}{L_3},\quad L_3=2d.
\end{equation}
A further ingredient is the real analytic Eisenstein series,
\begin{equation}
  \label{eq:38}
   f_2(\tau,\bar\tau)=\sum_{(n_{3},n_{4})\in \mathbb Z^2/(0,0)}
  \frac{\mathfrak{Im}(\tau)^2}{\abs{n_{4}+n_3\tau}^{4}},  
\end{equation}
which is invariant under modular transformations
\begin{equation}
  \label{eq:39}
  \tau'= \frac{a\tau +b}{c\tau +d}, \quad a,b,c,d\in \mathbb Z,\quad
  ad-bc=1. 
\end{equation}

Starting from the Lagrangian path integral formulation and its relation to the
appropriate zeta function, the partition function $\mathcal
Z(\tau,\bar\tau)=Z(\beta,\alpha)$ can be compactly written as
\begin{equation}
  \label{eq:40}
\ln{\mathcal Z}(\tau,\bar\tau)=\frac{1}{2\pi^{2}}
  \frac{L_1L_2}{L_3^{2}}
  \frac{f_{2}(\tau,\bar\tau)}{\mathfrak{Im}(\tau)}.
\end{equation}
Since $f_2(\tau,\bar\tau)$ is modular
invariant and $\mathfrak{Im}(\tau)$ transforms as
\begin{equation}
  \label{eq:41}
  \mathfrak{Im}(\tau')=\frac{\mathfrak{Im}(\tau)}{\abs{c\tau+d}^2},
\end{equation}
the partition function \eqref{eq:40} transforms as
\begin{equation}
  \label{eq:42}
  \ln{\mathcal Z}(\tau',\bar\tau')=\abs{c\tau+d}^2
     \ln{\mathcal Z}(\tau,\bar \tau).
\end{equation}
   
If the chemical potential vanishes, ${\alpha}=0$, $\tau=it$, with
$t=\frac{\beta}{L_3}$ and ${\rm Z}(t)=\mathcal Z(it,-it)=Z(\beta,0)$,
the result can be written in terms of an Epstein zeta function,
\begin{equation}
  \label{eq:59}
  \mathfrak Z(2;t^2,1)=\sideset{}{'}\sum_{n_{4},n_3}\frac{1}{(n_3^2t^2+n_{4}^2)^2},
\end{equation}
as 
\begin{equation}
  \label{eq:58}
  \ln{\rm Z}(t)=\frac{1}{2\pi^{2}}
  \frac{L_1L_2t}{L_3^{2}}\mathfrak Z(2;t^2,1).
\end{equation}
If $a=0=d$, $b=1=-c$, the modular transformation reduces to temperature
inversion $t\to \frac{1}{t}$ with
\begin{equation}
  \label{eq:60}
  \ln {\rm Z}(\frac{1}{t})=t^{2}\ln {\rm Z}(t).
\end{equation}

\subsection{Low and high temperature expansions from canonical approach}
\label{sec:low-high-temperature}

Let
\begin{equation}
  \label{eq:44}
  \xi(d)=\frac{\Gamma(\frac{d}{2})\zeta(d)}{\pi^{\frac{d}{2}}},\quad \xi(4)=\frac{\pi^2}{90},
  \quad \xi(3)=\frac{\zeta(3)}{2\pi}.
\end{equation}
The computation of the Casimir energy of the system yields the same result than
for electromagnetism and can be performed in several ways. The result is
\begin{equation}
  \label{eq:45}
  E_0(0)=-
      \xi(4)\frac{L_1L_2}{L_3^{3}}.
\end{equation}
The partition function can then be directly computed using a quantum statistical
approach in a Hilbert space basis that diagonalizes the Hamiltonian. The result
is
  \begin{multline}
    \label{eq:48}
     \ln{Z}(\beta,\alpha)=-\beta
  E_0+\xi(3)\frac{L_1L_2}{\beta^{2}}\\
 +2\frac{L_1L_2}{L_3^{\frac{3}{2}}\beta^{\frac{1}{2}}}
  \sideset{}{'}\sum_{n_3}\sum_{l\in\mathbb{N}^*}(\frac{\abs{n_3}}{l})^{\frac{3}{2}}
  K_{\frac{3}{2}}(2\pi 
      l\abs{n_3}\frac{\beta}{L_3})e^{2\pi i l n_3 \frac{ \alpha }{L_3}},
  \end{multline}
where $K_{\frac{3}{2}}(2\pi m \frac{\beta}{L_3})$ is a modified Bessel function
of the second kind. This result can also be written as 
\begin{multline}
  \label{eq:52}
  \ln{Z}(\beta,\alpha)=\frac{L_1L_2}{L_3^2}\Big[\xi(4)\frac{\beta}{L_3}
  +\xi(3)(\frac{L_3}{\beta})^2\\
 +2(\frac{L_3}{\beta})^{\frac{1}{2}}
  \sideset{}{'}\sum_{n_3}\sum_{l\in\mathbb{N}^*}(\frac{\abs{n_3}}{l})^{\frac{3}{2}}
  K_{\frac{3}{2}}(2\pi 
      l\abs{n_3}\frac{\beta}{L_3})e^{2\pi i l n_3 \frac{ \alpha }{L_3}}\Big].
\end{multline}

At low temperature/small distance, $\frac{\beta}{L_3}\gg 1$, the leading term in
the expansion to the partition function is directly related to the Casimir
energy. The leading correction is the contribution of the modes with spatial
frequencies $n_3=0$ and with normal ordered Hamiltonian. In this context, these
are the gravitons that propagate parallel to the plates, which is the sector of
the theory explicitly discussed in section \ref{sec:massl-scal-field}. It is
independent of the small dimension $L_3$ and coincides with the black body
result of electromagnetism in $2$ spatial dimensions or with a single massless
scalar field in $2$ spatial dimensions. On account of the asymptotic expansion
\begin{equation}
  \label{eq:47}
  K_\nu(z)=\sqrt{\frac{\pi}{2z}}e^{-z}(1+O(z^{-1})),
  \quad (\abs{{\rm arg}\ z}<\frac{3}{2}\pi), 
\end{equation}
for large $\abs{z}$, all other terms are exponentially suppressed.
It follows that in the low-temperature expansion of the entropy,
\begin{equation}
  \label{eq:49}
  S(\beta,\alpha)=(1-\beta\partial_\beta)\ln Z(\beta,\alpha)
  =3\xi(3)\frac{L_1L_2}{\beta^{2}}+\dots,
\end{equation}
the first term in \eqref{eq:48} proportional to the Casimir energy drops out
since it is linear in $\beta$. The dots denote exponentially suppressed terms.
The leading contribution now comes from the lower dimensional
electromagnetic/scalar field, i.e., the modes with $n_3=0$, and scales with the
area.

When expressed in terms of inverse temperature and chemical potential, the
generating set of transformations of the modular group become
\begin{equation}
  \label{eq:50}
  \tau'=\tau+1\iff \left\{\begin{array}{l} \beta'=\beta\\
                              \alpha'=\alpha+L_3
                            \end{array}
                          \right.\quad
                          \tau'=-\frac{1}{\tau}\iff  \left\{\begin{array}{l}
\frac{\beta'}{L_3}=\frac{L_3\beta}{\alpha^2+\beta^2}\\
 \frac{\alpha'}{L_3}= -\frac{L_3\alpha}{\alpha^2+\beta^2}
                            \end{array}
                          \right..   
\end{equation}
Under the first of these transformations, 
$\ln Z(\beta,\alpha)$ in \eqref{eq:48} is manifestly invariant, as required by
\eqref{eq:42} for $c=0,a=b=d=1$. For the second of these transformations, we get
\begin{equation}
  \ln Z(\beta',\alpha')=\frac{\alpha^2+\beta^2}{L_3^{2}}
    \ln Z(\beta,\alpha).\label{eq:85}
\end{equation}
When transposing and using the explicit expression in \eqref{eq:52} for the LHS,
this gives
\begin{multline}
  \label{eq:51}
  \ln Z(\beta,\alpha)=\frac{L_1L_2}{\alpha^2+\beta^2}
\Big[
  \xi(4)\frac{L_3\beta}{\alpha^2+\beta^2}
  +\xi(3)(\frac{\alpha^2+\beta^2}{L_3\beta})^{2}\\
  +2(\frac{\alpha^2+\beta^2}{L_3\beta})^{\frac{1}{2}}
   \sideset{}{'}\sum_{n_3}\sum_{l\in\mathbb N^*}(\frac{\abs{n_3}}{l})^{\frac{3}{2}}
  K_{\frac{3}{2}}(2\pi l\abs{n_3}\frac{L_3\beta}{\alpha^2+\beta^2})e^{-2\pi i l n_3
    \frac{L_3\alpha}{\alpha^2+\beta^2}}\Big]. 
\end{multline}
As a consequence, in the high temperature/large distance limit,
$\frac{\beta}{L_3}\ll 1$, the partition function is given by the first term of
the above equation. In the expression of the partition function in terms of the
real analytic Eisenstein \eqref{eq:40}, this term comes from the modes with
vanishing Matsubara frequencies $n_4=0$. For $\alpha=0$, this gives the standard
black body result \eqref{eq:74} when taking into account that the volume of the
system is $V=L_1L_2d$. More generally, the leading contributions are given by
the first two terms, while the others are exponentially suppressed. The high
temperature/large distance expansion of the entropy is
\begin{equation}
  \label{eq:43}
  S(\beta,\alpha)={L_1L_2}
  \Big[L_3\xi(4)\frac{4\beta^3}{(\alpha^2+\beta^2)^{3}}+
  \xi(3)\frac{3\alpha^2+\beta^2}{(L_3\beta)^{2}}+\dots\Big], 
\end{equation}
up to exponentially suppressed terms. When $\alpha=0$, this gives the black body
result corrected by a temperature-independent term,
\begin{equation}
  \label{eq:53}
  S(\beta,0)=8\xi(4)
  \frac{L_1L_2d}{\beta^{3}}+
  \frac{\xi(3)}{4}\frac{L_1L_2}{d^{2}}+\dots. 
\end{equation}
The leading term is directly related to the Casimir energy of the system, as
discussed in the context of the Cardy formula and modular invariance in higher
dimensions in \cite{Cappelli:1988vw,Cardy:1991kr,Shaghoulian:2015kta}.

\subsection{Complex trigonometric functions from method of images}
\label{sec:expr-terms-hyperb}

Both \eqref{eq:40} and \eqref{eq:52} can be shown to be equivalent to a single
series in terms of complex trigonometric functions, 
\begin{multline}
  \label{eq:46}
  \ln\mathcal Z(\tau,\bar\tau)= \frac{\pi^2L_1L_2\mathfrak{Im}(\tau)}{4L_3^2}
  \Big[\frac{2}{45}+\sum_{l\in \mathbb N^*}\big(
  \frac{i\cot{\pi l\tau }}{(\pi l\mathfrak{Im}(\tau))^3}\\
 -\frac{1}{(\pi l\mathfrak{Im}(\tau))^2\sin^2{\pi l\tau}}
  +{\rm c.c.}\big)\Big].
\end{multline}
If the chemical potential $\alpha$ vanishes, $\tau=it=i\frac{\beta}{L_3}$,
by using $\sin(ix)=i\sinh(x)$ and $\cot(ix)=-i\coth(x)$, we get 
\begin{equation}
  \label{eq:54}
  \ln{\rm Z}(t)= \frac{\pi^2L_1L_2t}{2L_3^2}
  \Big[\frac{1}{45}+\sum_{l\in \mathbb N^*}\big(
  \frac{\coth{\pi lt }}{(\pi l t)^3}
  +\frac{1}{(\pi l t)^2\sinh^2{\pi l t}}
  \big)\Big].
\end{equation}
This is the form under which the result usually appears in the Casimir
literature. It has originally been derived in \cite{Brown:1969na} by applying a
Sommerfeld-Watson transform to the Epstein zeta function \eqref{eq:59}, which in
turn had been obtained through the method of images.

\section{Discussion}
\label{sec:discussion-1}

As shown by the detailed Hamiltonian analysis, our version of perfectly
conducting boundary conditions for massless spin 2 fields are fully consistent
with gauge invariance. That the existence of such boundary conditions is
non-trivial has been mentioned for instance in \cite{Ambjorn:1981xw} (end of
section 4). It would be interesting to explore in detail how these boundary
conditions relate to those discussed recently in \cite{Witten:2018lgb}.

At the level of the free theory, an open question concerns ADM surface charges
for the set-up we are considering. This is relevant in relation to black holes
because in the Gibbons-Hawking treatment, these charges do play a crucial role.
It would be interesting to see whether they can be understood in terms of the
particles present in the spectrum, in analogy with what can be done for electric
charge in the case of the electromagnetic field between two perfectly conducting
plates. The other obvious question is to extend the analysis to more general
boundary geometries and to curved backgrounds.

With the boundary conditions that we have chosen, the Casimir force on the
``walls'' will be the same for gravitons than for photons. Trying to give a
meaning to this brings one back to the discussion in the introduction of what
type of wall that would be. More generally, in order to distinguish the result
for linearized gravity from that for electromagnetism and to begin to discuss
potential implications, one needs to consider interactions, and thus bring in
Newton's constant, in one way or another. We are looking forward to studying all
these questions in the near future.

\section*{Acknowledgements}
\label{sec:acknowledgements}

\addcontentsline{toc}{section}{Acknowledgments}

This work is supported by the F.R.S.-FNRS Belgium through conventions
FRFC PDR T.1025.14 and IISN 4.4503.15.

\appendix

\section{Photons in a Casimir box revisited}
\label{sec:reduced-phase-space-2}

In this appendix, we re-derive results on the scalar field formulation and
modular covariance in the electromagnetic finite temperature Casimir effect
\cite{Alessio:2020okv}. This is done in terms of suitable polarization vectors,
so that the construction may be generalized in a straightforward way to the
massless spin $2$ case.

\subsection{Boundary conditions and mode decomposition}
\label{sec:bound-cond-mode}

Consider coordinates $x^i$, $i=1,2,3$ in Euclidean space. The electric field is
$E^i=-\pi^i$, the magnetic field is $B^i=\epsilon^{ijk}\partial_j A_k$. The
starting point is the first order action
\begin{equation}
  \label{eq:17}
  S=\int \D x^0\Big[\int_V \D^3x\ \partial_0 A_i\pi^i -H+\int_V \D^3x\ A_0\partial_i \pi^i],\quad
  H=\frac{1}{2}\int_V \D^3x (\pi^i\pi_i+B^iB_i). 
\end{equation}

Let $a=1,2$, $i=a,3$, $V=L_1L_2d$ with $L_a$ large, $k_3=\frac{\pi n_3}{d}$,
$k_a=\frac{2\pi n_a}{L_a}$ (with no summation over $a$). Perfectly conducting
boundary conditions on parallel plates at $x^3=-\frac{d}{2}$ and
$x^3=\frac{d}{2}$ require $B^3=0$ and $E^a=0$ on the plates. Let
\begin{equation}
  \label{eq:14}
  \psi^H_{k}=\sqrt{\frac{2}{V}}e^{ik_a x^a}\sin k_3 x^3,\
    \psi^E_{k_a,0}=\frac{1}{\sqrt{V}}e^{ik_a x^a},\
    \psi^E_{k}=\sqrt{\frac{2}{V}}e^{ik_a x^a}\cos k_3 x^3,
\end{equation}
and let us use $V^i(x)$ for either of the canonically conjugate variables
$A^i(x)$ or $\pi^i(x)$. The boundary conditions are implemented through
\footnote{The factor $i$ in front of the expansion of $(A^a,\pi^a)$ is chosen
  for later convenience.}
\begin{equation}
  \label{eq:2}
    V^a(x)=i\sum_{n_a,n_3>0}V^{a}_{k}\psi^H_k,\quad
    V^3(x)=\sum_{n_a,n_3\geq 0}V^{3}_{k}\psi^E_{k},
  \end{equation}
where we take $V^a_{k_a,0}=0$. Reality and parity
conditions are
\begin{equation}
  \label{eq:3}
  V^{a}_{k_a,k_3}=-V^{*a}_{-k_a,k_3},\quad V^{3}_{k_a,k_3}=V^{*3}_{-k_a,k_3},
  \quad V^{a}_{k_a,k_3}=-V^{a}_{k_a,-k_3},\quad V^3_{k_a,k_3}=V^3_{k_a,-k_3}.
\end{equation}

\subsection{The particle}
\label{sec:part-resid-gauge}

The mode at $n_i=0$, $A_{3,0,0}=q,\pi^3_{0,0}=p$ is treated separately. It is
not affected by the constraints nor by proper gauge transformations. Its Poisson
brackets are canonical and its contribution to the Hamiltonian is that of a free
particle of unit mass,
\begin{equation}
H_{n_i=0}=\frac{1}{2}p^2\label{eq:20}.
\end{equation}

\subsection{Polarization vectors}
\label{sec:polarization-vectors}

Take now $\alpha=1,2$, and $A=(\alpha,\parallel)$. In Euclidean momentum space
$k^i$ (minus the origin) let $k=\sqrt{k_ik^i}$ and consider an orthonormal frame
${e_A}^i(k)$ built out of two vectors normal to $k^i$ and one vector parallel to
$k^i$,
\begin{equation}
  \label{eq:1}
  {e_\parallel}^i=\frac{k^i}{k},\quad {\epsilon^i}_{jm}{e_1}^j{e_2}^m={e_{\parallel}}^i,
  \quad {e_A}^i{e^B}_i=\delta^B_A,\quad {e_A}^i{e^B}_j=\delta^{B}_A, 
\end{equation}
so that the decomposition of a vector $v^i(k)$ in momentum space in this frame
is $V^i(k)=V^A(k) {e_A}^i$ with inverse $V^A(k)=V^i{e^A}_i$. 

The non-vanishing Poisson brackets for these modes are read off from the expansion of
$\int_{V}\D^3x\ \partial_0 A_i\pi^i$ and given by
\begin{equation}
  \label{eq:6}
  \{A_{A k},\pi^{*B}_{k'}\}=\delta_A^B\prod_{i=1}^3\delta_{n_i,n'_i}. 
\end{equation}

The contribution to the Hamiltonian from the $n_3> 0$ modes is given by
\begin{equation}
  \label{eq:5}
  H_{n_3\neq 0}=\frac{1}{2}\sum_{n_a,n_3>0}(\pi^{A}_k\pi^{*}_{k A}
  +k^2A^{\alpha}_kA^*_{k \alpha}).
\end{equation}
Let $k_\perp=\sqrt{k_a k^a}$. The contribution from the $n_3=0$ modes is
\begin{equation}
  \label{eq:7}
  H_{n_3=0}=\frac{1}{2}\sideset{}{'}\sum_{n_a}(\pi^{3}_{k_a,0}\pi^{*3}_{k_a,0}
  +k_\perp^2 A^{3}_{k_a,0}A^{*3}_{k_a,0}), 
\end{equation}
where the prime means that the mode with $n_i=0$ is omitted. In order to
implement the constraint $\partial_i\pi^i=0$ and/or the Coulomb gauge condition
$\partial^i A_i=0$,
\begin{equation}
  \label{eq:8}
  \partial_i V^i=-\sum_{n_a,n_3>0}kV^{\parallel}_k\psi^H_k=0\iff
  V^{\parallel}_{k}=0.
\end{equation}
Note that $V^a_{k_a,0}=0$ implies $V^\parallel_{k_a,0}=0$ and that, in order not
to introduce spurious variables, one needs to expand $A_0$ as
\begin{equation}
A_0=\sum_{n_a,n_3>0} A_{k,0}\psi^H_k\label{eq:19}.
\end{equation}
By variations with respect to $A_0$, respectively $A_{k,0}$, one may then solve
the constraints, or equivalently $\pi^{\parallel}_{k}=0$ in the action, without
the need to impose the Coulomb or any other gauge condition. As a consequence,
in the first term in \eqref{eq:5}, one may limit the sum over $A$ to one over
$\alpha$. The same is true for the kinetic term that gives rise to the canonical
Poisson brackets. Because proper gauge transformations with gauge parameters
satisfying Dirichlet conditions correspond to arbitrary shifts of
$A^\parallel_k$, it follows that gauge invariant quantities reduced to the
constraint surface do not depend on the variables
$A^\parallel_k,\pi^\parallel_k$.

The electric charge observable is gauge invariant and given by 
\begin{equation}
  Q=\int_{x^3} \D^2x\ \pi^3=-L_1L_2 \sum_{n_3>0}\pi^3_{0,k_3}\psi^E_{0,k_3}
  -\sqrt{\frac{L_1L_2}{d}}p,
  \label{eq:21}
\end{equation}
where integration is at fixed $x_3\in ]-\frac{d}{2},\frac{d}{2}[$. Since
$\pi^3_{0,k_3}=\pi^\parallel_{0,k_3}$, the observable reduces to the last term
only on the constraint surface and is $x_3$ independent.

\subsection{Adapted polarization vectors}
\label{sec:adap-polar-vect} 

Consider now the following choice of ${e_{\alpha}}^i$ (see e.g.~\cite{Bordag:2009zzd}),
\begin{equation}
  \label{eq:9}
  {e_{H}}^i=\frac{1}{k_{\perp}}\begin{pmatrix}k_2\\-k_1\\0
\end{pmatrix}
  \qquad {e_{E}}^i=\frac{1}{k_{\perp}k}\begin{pmatrix} k_1k_3\\k_2k_3\\-k^2_{\perp}
  \end{pmatrix},
\end{equation}
so that the components $V^{H}$, $V^E$, $V^\parallel$ are given by
\begin{equation}
  \label{eq:10}
  V^{H}_{k_a,k_3}=\frac{\epsilon^{ab}k_b}{k_{\perp}}V^a_{k_a,k_3}, \quad
  V^E_{k_a,k_3}=\frac{k_ak_3}{k_{\perp}k}V^{a}_{k_a,k_3}-\frac{k_{\perp}}{k}
  V^{3}_{k_a,k_3},\quad V^\parallel_{k_a,k_3}=\frac{k_i}{k}V^i_{k_a,k_3},
\end{equation}
with inverse relations
\begin{equation}
  \label{eq:12}
  V^{a}_{k_a,k_3}=\frac{\epsilon^{ab}k_b}{k_{\perp}}V^{H}_{k_a,k_3}
  +\frac{k^ak_3}{k_{\perp}k}V^{E}_{k_a,k_3}+k^aV^{\parallel}_{k_a,k_3},\quad 
V^{3}_{k_a,k_3}=-\frac{ k_{\perp}}{k}V^{E}_{k_a,k_3}+k_3V^{\parallel}_{k_a,k_3}.  
\end{equation}
Note in particular that $V^3_{k_a,0}=-V^{E}_{k_a,0}$. Reality and parity
conditions become
\begin{equation}
  \label{eq:11}
  V^{H}_{k_a,k_3}=V^{*H}_{-k_a,k_3},\quad V^{E}_{k_a,k_3}=V^{*E}_{-k_a,k_3},\quad
  V^{H}_{k_a,-k_3}=-V^{H}_{k_a,k_3},\quad V^{E}_{k_a,k_3}=V^{E}_{k_a,-k_3}.
\end{equation}
Let $\lambda=(H,E)$. Oscillator variables are defined as
\begin{equation}
  \label{eq:13}
\begin{split}
  &a^{\lambda}_{k_a,k_3}=\sqrt{\frac{k}{2}}(A^{\lambda}_{k_a,k_3}
  +\frac{i}{k}\pi^{\lambda}_{k_a,k_3}),\\ &A^{\lambda}_{k_a,k_3}
  =\frac{1}{\sqrt{2k}}(a^{\lambda}_{k_a,k_3}+a^{*\lambda}_{-k_a,k_3}),\quad
  \pi^{\lambda}_{k_a,k_3}
  =-i\sqrt{\frac{k}{2}}(a^{\lambda}_{k_a,k_3}-a^{*\lambda}_{-k_a,k_3}),
\end{split}
\end{equation}
with the understanding that $a^H_{k_a,0}=0$. Their non-vanishing Poisson
brackets are 
\begin{equation}
  \label{eq:23}
  \{a^\lambda_{k},a^{*\lambda'}_{k'}\}=-i\delta^{\lambda,\lambda'}
  \prod_{i=1}^3\delta_{n_i,n'_i}. 
\end{equation}
In terms of these oscillators, we
have
\begin{equation}
  \label{eq:4}
  \begin{split}
    A^a&=i\sum_{n_a,n_3>0}\Big[\frac{\epsilon^{ab}k_b}{\sqrt{2k}k_{\perp}}
    (a^H_{k}\psi^H_k-{\rm
      c.c.}) +\frac{k^ak_3}{\sqrt{2k}k_{\perp} k}(a^E_{k}\psi_k^H-{\rm c.c.})
    +k^aA^{\parallel}_{k}\psi_k^H\Big],\\
    \pi^a&=\sum_{n_a,n_3>0}\Big[\frac{\sqrt{k}\epsilon^{ab}k_b}{\sqrt{2}k_{\perp}}
    (a^H_{k}\psi^H_k+{\rm c.c.}) +\frac{k^ak_3}{\sqrt{2k}k_{\perp}
      }(a^E_{k}\psi_k^H+{\rm c.c.}) +ik^a\pi^{\parallel}_{k}\psi_k^H\Big],\\
      A^3&=\sideset{}{'}\sum_{n_a,n_3\geq 0}\Big[-\frac{k_{\perp}}{\sqrt{2k}k}
      (a^E_{k}\psi_k^E+{\rm
      c.c.})+ k_3A^{\parallel}_{k}\psi^E_k\Big]+\frac{1}{\sqrt{V}}q,\\
    \pi^3&=\sideset{}{'}\sum_{n_a,n_3\geq 0}\Big[i\frac{k_{\perp}}{\sqrt{2k}}
    (a^E_{k}\psi_k^E-{\rm
      c.c.})+ k_3\pi^{\parallel}_{k}\psi^E_k\Big]+\frac{1}{\sqrt{V}}p.
\end{split}
\end{equation}
On the constraint surface, the full Hamiltonian is given by
\begin{equation}
  \label{eq:22}
    H=\sideset{}{'}\sum_{\lambda,n_a,n_3\geq 0}
    \frac{1}{2} (\pi^\lambda_{k}\pi^{*\lambda}_k+
    k^2A_k^\lambda A^{*\lambda}_k)+\frac{1}{2}p^2
    =\sideset{}{'}\sum_{\lambda,n_a,n_3\geq
      0}\frac{k}{2} (a^\lambda_ka^{*\lambda}_k+ a^{*\lambda}_ka^\lambda_k) +\frac{1}{2}p^2.
\end{equation}
Besides the electric charge operator
\begin{equation}
  \label{eq:24}
  Q=-\sqrt{\frac{L_1L_2}{d}}p,
\end{equation}
another observable of interest to us is directly related to spin angular
momentum of light. In reduced phase space, the latter is given by
\begin{equation}
  \label{eq:25}
  J_i=\int_V \D^3x\ \epsilon_{ijk} A_\perp^j\pi_\perp^k=\int_V \D^3x\ \vec \pi_\perp
  \cdot (\vec e_i\times \vec A_\perp), 
\end{equation}
where $V_i^\perp$ correspond to the vectors in \eqref{eq:4} with
$V^\parallel_k=0=q=p$.
When using that 
\begin{equation}
{e_H}^i{e_E}^j-{e_E}^i{e_H}^j=\epsilon^{ijl}\frac{k_l}{k}\label{eq:106},
\end{equation}
it follows that, in terms of modes,
\begin{equation}
  \label{eq:27}
    J_3=\sum_{n_a,n_3>0}\frac{k_3}{k}(A^H_k\pi^{*E}_k-A^E_k\pi^{*H}_k)
    = \sum_{n_a,n_3>0}\frac{ik_3}{k}(a^{*H}_ka^{E}_k-a^{*E}_ka^{H}_k), 
\end{equation}
More precisely, because we are interested in a modular covariant partition
function, we are interested in the observable
\begin{equation}
  \label{eq:26}
  P^S_3=-\int_V \D^3x\  \partial_3 B^i \frac{1}{\sqrt{-\triangle}}\pi_i,  
\end{equation}
whose mode expansion can be shown to differ from that of $J_3$ by multiplying
each term in momentum space by $k$,
\begin{equation}
  \label{eq:28}
    P^S_3=\sum_{n_a,n_3>0}{k_3}(A^H_k\pi^{*E}_k-A^E_k\pi^{*H}_k)
    = \sum_{n_a,n_3>0}{ik_3}(a^{*H}_ka^{E}_k-a^{*E}_ka^{H}_k). 
\end{equation}

\subsection{Bromwich-Borgnis fields}
\label{sec:borgnis-fields}

Consider now the real fields
\begin{equation}
  \label{eq:15}
  \begin{split}
    \phi^{H}&=\sum_{n_a,n_3>0}\frac{1}{\sqrt{2k}k_{\perp}}(a^H_{k}\psi^H_k+{\rm c.c.}),\quad \pi^H
    =-i\sum_{n_a,n_3>0}\frac{\sqrt{k}}{\sqrt{2}k_{\perp}}(a^H_{k}\psi^H_k-{\rm c.c.})\\
  \phi^{E}&=-\sideset{}{'}\sum_{n_a,n_3\geq 0}\frac{1}{\sqrt{2k}kk_{\perp}}(a^E_{k}\psi^E_k
  +{\rm c.c.}),\quad \pi^{E}=i\sideset{}{'}\sum_{n_a,n_3\geq 0}\frac{1}{\sqrt{2k}k_{\perp}}(a^E_{k}\psi^E_k
  -{\rm c.c.})\\
  \phi^G&=\sum_{n_a,n_3>0}\frac{1}{2}(A^\parallel_k\psi^H_k+{\rm c.c.}),\quad
  \pi^G=\sum_{n_a,n_3>0}\frac{1}{2}(\pi^\parallel_k\psi^H_k+{\rm c.c.}).
\end{split}
\end{equation}
Let $\Lambda=(H,E,G)$ and $\varphi^\Lambda$ stand for either $\phi^\Lambda$ or
$\pi^\Lambda$. These fields all satisfy the Helmholtz equation
\begin{equation} 
  \label{eq:16}
  (\triangle+\partial^2_3)\varphi^\Lambda=0,
\end{equation}
with $\varphi^H,\varphi^G$ satisfying Dirichlet conditions while $\varphi^E$
satisfies Neumann conditions. In these terms,
\begin{equation}
  \label{eq:18}
  \begin{split}
    A^a&=\epsilon^{ab}\partial_b\phi^H+\partial^a\partial_3\phi^E+\partial^a\phi^G,\quad
    A^3=(-\triangle+\partial^2_3)\phi^E+\partial^3\phi^G+\frac{1}{\sqrt V}q,\\
    \pi^a&=\epsilon^{ab}\partial_b\pi^H+\partial^a\partial_3\pi^E+\partial^a\pi^G,\quad
    \pi^3=(-\triangle+\partial^2_3)\pi^E+\partial^3\pi^G+\frac{1}{\sqrt V}p,\\
    B^a&=\epsilon^{ab}\partial_b(-\triangle)\phi^E+\partial^a\partial_3\phi^H,\quad
    B^3=(-\triangle+\partial_3^2)\phi^H. 
  \end{split}
\end{equation}
On the constraint surface where $\pi^G=0$, one recovers the construction of
\cite{doi:10.1080/14786440708635935,doi:10.1002/andp.19394270408} (see also
\cite{Phillips1962} section 32 and
\cite{DeWitt:1975ys,Deutsch1979,Plunien:1986ca} for related more modern
discussions).

\subsection{Single scalar field formulation}
\label{sec:single-scalar-field}

In order to streamline the computation of the partition function and to discuss
modular properties, it is useful to go one step further and introduce a
formulation with a single scalar field on $z\in [-d,d]$ with periodic boundary
conditions \cite{Alessio:2020okv}.

The variables defined by 
\begin{equation}
  \label{eq:29}
  \begin{split}
    &\phi_{k_a,k_3}=\frac{A^E_{k_a,k_3}-iA^{H}_{k_a,k_3}}{\sqrt{2}},\quad
    \pi_{k_a,k_3}=\frac{\pi^E_{k_a,k_3}-i\pi^{H}_{k_a,k_3}}{\sqrt{2}},\\
    &\phi_{k_a,0}=A^E_{k_a,0},\quad \pi_{k_a,0}=\pi^E_{k_a,0},
  \end{split}
\end{equation}
satisfy the reality conditions
\begin{equation}
  \label{eq:30}
  \phi_{-k_a,0}^*=\phi_{k_a,0},\quad \phi_{-k_a,-k_3}^*=\phi_{k_a,k_3},\quad
  \pi_{-k_a,0}^*=\pi_{k_a,0},\quad \pi_{-k_a,-k_3}^*=\pi_{k_a,k_3}.
\end{equation}
The first expression for the Hamiltonian in \eqref{eq:22} can then be written as
\begin{equation}
  \label{eq:31}
  H=\sum_{n_i}\frac{1}{2}(\pi_{k}\pi^*_{k}+k^2 \phi_{k}\phi^*_{k}),
\end{equation}
with the understanding that $\pi_{0,0}=p$, $\phi_{0,0}=q$ and the sum goes over
all $n_i\in\mathbb Z$. In terms of
appropriate oscillators, defined for $n_i\neq 0$,
\begin{equation}
  \label{eq:32}
  a_{k_a,k_3}=\sqrt{\frac{k}{2}}(\phi_{k_a,k_3}+\frac{i}{k}\pi_{k_a,k_3})
  =\left\{\begin{array}{l} \frac{a^{E}_{k_a,k_3}-ia^{H}_{k_a,k_3}}{\sqrt{2}},\quad n_3\neq 0,\\
a^{E}_{k_a,0}, \quad n_3\neq 0
         \end{array}\right.,
\end{equation}
the Hamiltonian becomes
\begin{equation}
  \label{eq:33}
  H=\sideset{}{'}\sum_{n_i}\frac{k}{2}(a^{*}_{k}a_{k}+a_{k}a^{*}_{k})+\frac{1}{2} p^2. 
\end{equation}
The above are the mode decomposition, oscillators and Hamiltonian of a single
real scalar field $\phi$ and its momentum $\pi $ in a volume $V'=L_1L_2L_3$
where $L_3=2d$ with periodic boundary conditions in all directions and, in
particular, with periodicity $2d$ in the $x^3$ direction.
\begin{equation}
  \label{eq:34}
  \begin{split}
    &\phi=\frac{1}{\sqrt{V'}}\sum_{n_i}e^{ik_jx^j}\phi_{k_i}
    =\frac{1}{\sqrt{V'}}\Big(\sideset{}{'}\sum_{n_i}\frac{1}{\sqrt{2k}}
    [a_{k_i}e^{ik_jx^j}+a^{*}_{k_i}e^{-ik_jx^j}]+q\Big), \\
    &\pi=\frac{1}{\sqrt{V'}}\sum_{n_i}e^{ik_jx^j}\pi_{k_i}=
   \frac{1}{\sqrt{V'}}\Big(-i\sideset{}{'}\sum_{n_i}\sqrt{\frac{k}{2}}
      [a_{k_i}e^{ik_jx^j}-a^{*}_{k_i}e^{-ik_jx^j}]+p\Big),
  \end{split}
\end{equation}
for which the Hamiltonian is
\begin{equation}
  \label{eq:114}
  H=\frac{1}{2}\int_{V'}\D^3x\ (\pi^2+\partial_i\phi\partial^i\phi).
\end{equation}
By design of $P_3^S$ in \eqref{eq:26}, the mode expansion of linear momentum
along $x^3$ of the scalar field,
\begin{equation}
  \label{eq:115}
  P^S_3=-\int_{V'}\D^3x\ \pi\partial_3\phi=\sum_{n_i}\ k_3a^*_ka_k, 
\end{equation}
can be easily shown to agree with \eqref{eq:28} when using \eqref{eq:32}.

\subsection{Partition function of photons}
\label{sec:extend-part-funct-1}

For the computation of the partition function for photons in a Casimir box,
\begin{equation}
  \label{eq:35}
  Z(\beta,\alpha,\mu)={\rm Tr} e^{-\beta (\hat H-\mu\hat Q)+\alpha \hat P^S_3},
\end{equation}
one may now use the scalar field formulation to show that
\begin{equation}
  \label{eq:36}
  \ln Z(\beta,\alpha,\mu)=\frac{L_1L_2}{L_3}\beta\mu^2
  +\ln \mathcal Z(\tau,\bar\tau). 
\end{equation}
The first terms comes from the particle and reproduces the Gibbons-Hawking
contribution to the planar capacitor \cite{Barnich:2018zdg,Barnich2019}. It is a
classical contribution on top of the standard contribution corresponding to the
Casimir energy at zero temperature. The second term corresponds to the modular
covariant partition function of a massless scalar field on $\mathbb
S^1_{2d}\times \mathbb R^2$. Several equivalent expressions are given in the
main text in section \ref{sec:extend-part-funct}.

\section{Linearized gravity in empty space}
\label{sec:reduced-phase-space}

\subsection{Hamiltonian formulation of Pauli-Fierz theory}
\label{sec:spac-form}

The Hamiltonian formulation of massless spin 2 fields may be obtained by
linearizing the ADM formulation of full general relativity around flat space,
with the result
\begin{equation}
  \label{eq:55}
  S_{\rm PF}[h_{ij},\pi^{ij},n_i,n]=\int \D x^0\Big[\int
  \D^3x\, \big(\pi^{ij}\partial_0 h_{ij}-n^i\mathcal H_i-n\mathcal H_\perp \big)-H_{\rm PF}\Big],
\end{equation}
where
\begin{multline}
  \label{eq:56}
  H_{\rm PF}[h_{ij},\pi^{ij}]=\int
  \D^3x\big(\pi^{ij}\pi_{ij}-\frac{1}{2}\pi^2+\frac{1}{4}\partial^lh^{ij}
  \partial_lh_{ij}-\\
  -\frac{1}{2}\partial_i h^{ij}\partial^l h_{lj}+\frac{1}{2}\partial^i
  h\partial^j h_{ij}-\frac{1}{4}\partial^i h\partial_i h\big),
\end{multline}
and
\begin{equation}
  \label{eq:57}
  \mathcal H_i=-2\partial^j\pi_{ij},\quad
  \mathcal H_\perp=\Delta h-\partial^i\partial^j h_{ij}. 
\end{equation}
Indices are lowered and raised with the flat space metric $\delta_{ij}$ and its
inverse, $h={h^i}_i$, $\pi={\pi^i}_i$ and $\Delta =\partial_i\partial^i$ is the
Laplacian in flat space. The linearized $4$ metric is reconstructed using
$h_{00}=-2n$ and $h_{0i}= n_i$.

One may now discuss the decomposition of symmetric rank two tensors
$\phi_{ij},\pi^{ij}$ into transverse trace-less, trace and longitudinal parts
\cite{Arnowitt:1962aa,Deser1967} either in real space or in reciprocal space.
For the purpose of computing a partition function in physical Fock space, we
will do the latter. In what follows, we thus consider the Fourier transforms of
the fields
\begin{equation}
\phi(k)=\frac{1}{(2\pi)^{\frac{3}{2}}}\int \D^3x\ e^{-i
  k_jx^j}\phi(x)\label{eq:62}.
\end{equation}

\subsection{Polarization tensors}
\label{sec:fourier-space}

Let ${e_A}^i$, $A=(\alpha,\parallel)$, $\alpha=1,2$ be the orthonormal frame for
vectors as in \eqref{eq:1}. Let $\Xi=(TTs,T,LT\alpha,LL)$, with $s=(+,\times)$.
When $k_i\neq 0$, an orthonormal basis for symmetric tensors ${e_{\Xi}}^{ij}$ is constructed as
follows:
\begin{equation}
  \label{eq:63}
  \begin{split}
  & {e_{TT+}}^{ij} =\frac{1}{\sqrt 2}({e_1}^i{e_1}^j-{e_2}^i{e_2}^j),\quad
  {e_{TT \times}}^{ij}=\frac{1}{\sqrt 2}({e_1}^i{e_2}^j+{e_2}^i{e_1}^j),\\
  & k_i\ {e_{TTs}}^{ij} =0,\quad \delta_{ij}\ {e_{TTs}}^{ij}=0,
\end{split}
\end{equation}
\begin{equation}
  \label{eq:64}
  {e_T}^{ij}=\frac{1}{\sqrt 2}(\delta^{ij}-{e_\parallel}^i{e_{\parallel}}^j),
  \quad k_i\ {e_T}^{ij}=0,\quad \delta_{ij}\ {e_T}^{ij}=\sqrt 2,  
\end{equation}
\begin{equation}
  \label{eq:65}
  {e_{LT\alpha}}^{ij}=\frac{1}{\sqrt 2}({e_\parallel}^i{e_\alpha}^j
  +{e_\alpha}^i{e_\parallel}^j),\quad k_i {e_{LT\alpha}}^{ij}=\frac{k}{\sqrt 2}{e_\alpha}^j,\quad
  \delta_{ij}{e_{LT\alpha}}^{ij}=0,
\end{equation}
\begin{equation}
  \label{eq:66}
  {e_{LL}}^{ij}={e_\parallel}^i{e_\parallel}^j,\quad k_i {e_{LL}}^{ij}= k^j,\quad \delta_{ij}{e_{LL}}^{ij}=1. 
\end{equation}
This basis is orthonormal,
\begin{equation}
  \label{eq:67}
  {e_{\Xi}}^{ij}{e^\Gamma}_{ij}=\delta^\Gamma_\Xi,\quad {e_{\Xi}}^{ij} {e^\Xi}_{mn}
  =\frac{1}{2}(\delta^i_m\delta^j_n+\delta^i_n\delta^j_m),
\end{equation}
so that the Fourier components of $h_{mn},\pi^{mn}$ may be decomposed as 
\begin{equation}
  \label{eq:68}
  h_{ij}=h_\Xi {e^\Xi}_{ij},\quad h_\Xi= h_{ij}{e_\Xi}^{ij},\quad
  \pi^{ij}=\pi^\Xi {e_\Xi}^{ij},\quad  \pi^\Xi=\pi^{ij} {e_{ij}}^\Xi. 
\end{equation}

With this decomposition, the kinetic term becomes
\begin{equation}
  \label{eq:69}
  \int\D^3x\ \dot h_{ij}\pi^{ij}=\int \D^3k\ \dot h_\Xi\pi^{*\Xi}, 
\end{equation}
with the usual reality conditions, so that the non-vanishing Poisson are
\begin{equation}
  \label{eq:70}
  \{h_\Xi(k),\pi^{*\Gamma}(k')\}=\delta_\Xi^\Gamma\delta^3(k,k'). 
\end{equation}
Decomposing the Fourier components of the shift as $n^i(k)=n^A(k) {e_A}^i$, the
constraints in the action principle become 
\begin{equation}
  \label{eq:71}
  \int \D^3x\, \big(n^i\mathcal H_i+n\mathcal H \big)
  =\int \D^3k\ (2ik)(n^\parallel\pi^*_{LL}+n^\alpha \frac{1}{\sqrt 2}\pi^*_{LT\alpha}
  -n k^2\sqrt{2}h^*_T).
\end{equation}
The constraint surface is thus determined by $\pi^{LL}=0=\pi^{LT\alpha}=h_T$. On
the constraint surface, the first order action reduces to
\begin{equation}
  \label{eq:61}
  \begin{split}
    S_{\rm PF}\approx \int \D x^0\Big[\int \D^3k\ \partial_0 h_{TTs}\pi^{*TTs}
    -H_{\rm PF}\Big],\\ H_{\rm PF}\approx \int \D^3k\
    (\pi_{TTs}\pi^{*TTs}+\frac{1}{4} \pi_{TTs}\pi^{*TTs}),
\end{split}
\end{equation}
without the need to fix a specific gauge.

Defining the oscillator variables, 
\begin{equation}
a_{TTs}=\frac{1}{2}\sqrt{k} h_{TTs}+i\frac{\pi^{TTs}}{\sqrt k}\label{eq:72}, 
\end{equation}
up to integrations by parts in $\partial_0$, the first order action becomes 
\begin{equation}
  \label{eq:73}
  \begin{split}
  S_{\rm PF} &\approx \int \D x^0\Big[\int \D^3k\ \frac{1}{2i}(\partial_0
  a^*_{TTs}a^{TTs}-a^*_{TTs}\partial_0 a^{TTs})-H_{\rm PF}\Big],\\
  H_{\rm PF} &\approx \int \D^3k\ k\, a^*_{TTs}a^{TTs}.
\end{split}
\end{equation}
In the large volume limit, the partition function $Z(\beta)={\rm Tr} e^{-\beta
  :\hat H_{\rm PF}:}$, involving the normal ordered Hamiltonian, is then
computed as usual by first putting the system in a box, for instance with
periodic boundary conditions. This box is subsequently taken to be large so that
one may replace sums by integrals, up to a volume factor. This yields the
standard black body result,
\begin{equation}
  \label{eq:74}
  \ln Z(\beta) =2\frac{\pi^2}{90}\frac{V}{\beta^3}. 
\end{equation}

\section{Generalized curl and projection operators on modes}
\label{sec:gener-curl-proj}

Let us consider
\begin{equation}
  \label{eq:125}
  \begin{split}
    & (\hat{\mathcal O} T_k)^{ij}=k (\vec{e_{\parallel}}\times T_k)^{ij}
    =\frac{k}{2}(\epsilon^{i}_{lm}{e_\parallel}^l{T_k}^{mj}+\epsilon^{j}_{lm}{e_\parallel}^l{T_k}^{im}),\\
    & (\hat{\mathcal O} 
    {e_{\Xi}})^{ij}T^\Xi_k=k({e_{TT\times}}^{ij}T_k^{TT+}-{e_{TT+}}^{ij}T_k^{TT\times}
    +\frac{1}{2}{e_{LTE}}^{ij}T_k^{LTH}
    -\frac{1}{2}{e_{LTH}}^{ij}T_k^{LTE}),
  \end{split}
\end{equation}
and
\begin{equation}
  \begin{split}
    &(\hat{\mathcal P} T_k)^{ij}=k^2(T^{ij}_k-{e_i}^\parallel{e_l}^\parallel
    {T_k}^l_j
    -{e_j}^\parallel{e_l}^\parallel {T_k}^l_i),\\
    &(\hat{\mathcal P}
    {e_{\Xi}})^{ij}T^\Xi_k=k^2({e_{TT+}}^{ij}T_k^{TT+}+{e_{TT\times}}^{ij}T_k^{TT\times}
    +{e_{T}}^{ij}T_k^{T}-{e_{LL}}^{ij}T_k^{LL}).\label{eq:127}
  \end{split}
\end{equation}
In terms of the mode expansion for the Casimir box, the actions of $\mathcal O$
in \eqref{eq:108} and $\mathcal P$ in \eqref{eq:117} are explicitly given by
\begin{equation}
  \label{eq:126}
  \begin{split}
     & (\mathcal O T)^{ab}=\sideset{}{'}
  \sum_{n_a,n_3\geq 0}(\hat{\mathcal O} T_k)^{ab}\psi_k^E=\sideset{}{'}
  \sum_{n_a,n_3\geq 0}(\hat{\mathcal O} 
  {e_{\Xi}})^{ab}T^\Xi_k\psi^E_k,\\
  &(\mathcal O T)^{a3}=\sum_{n_a,n_3> 0}i(\hat{\mathcal O} T_k)^{a3}\psi_k^H=
  \sum_{n_a,n_3 >0}i(\hat{\mathcal O} 
  {e_{\Xi}})^{a3}T^\Xi_k\psi^H_k,\\
  & (\mathcal O T)^{33}=\sideset{}{'}
  \sum_{n_a,n_3\geq 0}(\hat{\mathcal O} T_k)^{33}\psi_k^E=\sideset{}{'}
  \sum_{n_a,n_3\geq 0}(\hat{\mathcal O} 
  {e_{\Xi}})^{33}T^\Xi_k\psi^E_k,\\
  \end{split}
\end{equation}
and
\begin{equation}
  \label{eq:128}
  \begin{split}
    & (\mathcal P T)^{ab}= 
    \sum_{n_a,n_3> 0}(\hat{\mathcal P} T_k)^{ab}\psi_k^H=\sum_{n_a,n_3> 0}(\hat{\mathcal P}
    {e_{\Xi}})^{ij}T^\Xi_k\psi^H_k,\\
    & (\mathcal P T)^{a3}= 
    \sideset{}{'}
  \sum_{n_a,n_3\geq 0}(-i)(\hat{\mathcal P} T_k)^{a3}\psi_k^E=\sideset{}{'}
  \sum_{n_a,n_3\geq 0}(-i)(\hat{\mathcal P}
    {e_{\Xi}})^{a3}T^\Xi_k\psi^E_k,\\
    & (\mathcal P T)^{33}= 
    \sum_{n_a,n_3> 0}(\hat{\mathcal P} T_k)^{33}\psi_k^H=\sum_{n_a,n_3> 0}(\hat{\mathcal P}
    {e_{\Xi}})^{33}T^\Xi_k\psi^H_k.
  \end{split}
\end{equation}

\vfill
\pagebreak

\addcontentsline{toc}{section}{References}


\providecommand{\href}[2]{#2}\begingroup\raggedright\endgroup

\end{document}